# DESIGN AND SIMULATION OF IOTA - A NOVEL CONCEPT OF INTEGRABLE OPTICS TEST ACCELERATOR

S. Nagaitsev, A. Valishev, Fermilab, Batavia, IL 60510, U.S.A.
D. Shatilov, BINP SB RAS, Novosibirsk, 630090, Russia
V. Danilov, ORNL, Oak Ridge, TN 37831, U.S.A.

*Abstract*

The use of nonlinear lattices with large betatron tune spreads can increase instability and space charge thresholds due to improved Landau damping. Unfortunately, the majority of nonlinear accelerator lattices turn out to be nonintegrable, producing chaotic motion and a complex network of stable and unstable resonances. Recent advances in finding the integrable nonlinear accelerator lattices have led to a proposal to construct at Fermilab a test accelerator with strong nonlinear focusing which avoids resonances and chaotic particle motion. This presentation will outline the main challenges, theoretical design solutions and construction status of the Integrable Optics Test Accelerator (IOTA) underway at Fermilab.

## INTRODUCTION

In 1952 the groundbreaking work [1] of Courant, Livingston and Snyder has revolutionized the accelerators by introducing the principle of alternating-gradient (also known as strong) beam focusing. They have discovered that an arrangement of alternating (focusing-defocusing) quadrupole and dipole magnets can keep the charged particle focused on average. Fundamentally, this is possible because the particle transverse motion, despite being time dependent, possesses two invariants of motion, known now as Courant-Snyder invariants. The existence of motion invariants always simplifies the motion, introducing constraints on particle dynamics. If two commuting integrals of motion exist, the particle's motion (if finite) could be reduced to motion on tori in 4D phase space.

All present accelerators with strong focusing have the following property: particle transverse oscillation (betatron) frequency is by design independent of particle amplitude. The particle dynamics can be understood by introducing the so-called normalized phase-space coordinates:

$$z_N = \frac{z}{\sqrt{\beta(s)}}, \quad p_N = p\sqrt{\beta(s)} - \frac{\beta'(s)z}{2\sqrt{\beta(s)}}, \quad (1)$$

where $z$ stands for either $x$ or $y$ particle coordinates, $p$ is similarly either $p_x$ or $p_y$, and $\beta(s)$ is either the horizontal or vertical beta-function. In these normalized coordinates the particle motion is identical to that of a linear oscillator:

$$\frac{d^2 z_n}{d\psi^2} + \omega^2 z_n = 0, \quad (2)$$

where $\psi$ is the new "time", which is the betatron phase

$$\psi' = \frac{1}{\beta(s)}. \quad (3)$$

In this paper we will present several examples of focusing systems, which are nonlinear by design, with the particle frequency being dependent on amplitude, yet stable and integrable. The benefits of such a system are two-fold. First, the increased betatron frequency spread provides improved Landau damping. Second, a nonlinear system is more stable to perturbations than a linear one [2].

There are only a handful of nonlinear systems with one or two analytic integrals of motion for accelerators. The reason is that the nonlinear systems with analytic invariants are very rare in a vast sea of nonlinear systems. In addition, for the focusing elements one has to use only the electromagnetic fields obeying the Maxwell equations. This, in turn, drastically reduces the variety of systems with invariants for practical use.

In the next Section we describe the design choice for the Integrable Optics Test Accelerator Ring (IOTA) aimed at testing nonlinear focusing systems and other new accelerator ideas as well. In the following we present what we think are good candidates for first tests of nonlinear systems producing large frequency spreads and low particle loss.

## IOTA DESIGN

The IOTA ring was designed for the proof-of-principle experiment of the nonlinear integrable optics idea [3] at the ASTA facility [4]. The initial version of the ring design described in [5] was comprised of four periodic cells and had full 8-fold mirror symmetry. It was later identified that it is desirable to accommodate more experiments, such as the nonlinear focusing with electron beam lens [6] and optical stochastic cooling. These options demanded incorporation of a 5 m-long straight section. The experimental hall dimensions allow to accommodate the stretched ring (Fig. 1) keeping the four 2 m nonlinear magnet insertions and the e- beam energy of 150 MeV. Table 1 summarizes the main parameters of IOTA. The straight section opposite to the long experimental insertion will be used for injection, RF cavity and instrumentation. The ring lattice is comprised of 50 conventional water-cooled quadrupole and 8 dipole magnets. The beam pipe aperture is 50 mm.

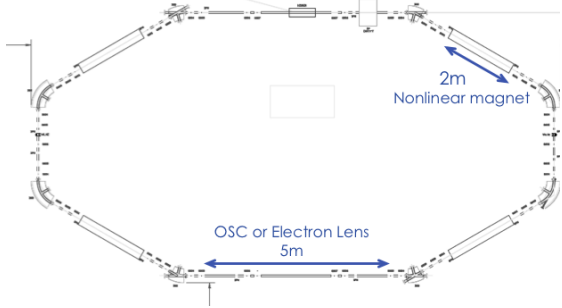

Figure 1: Layout of the IOTA ring.

Table 1: Summary the main parameters of IOTA

| Parameter | Value |
| --- | --- |
| Nominal beam energy | 150 MeV($\gamma$=295) |
| Nominal beam intensity | $1\times10^9$ (single bunch) |
| Circumference | 38.7 m |
| Bending field | 0.7 T |
| Beam pipe aperture | 50 mm dia. |
| Maximum $\beta$-function | 3 ÷ 9 m |
| Momentum compaction | 0.015 ÷ 0.1 |
| Betatron tune | 3.5 ÷ 7.2 |
| Natural chromaticity | -5 ÷ -15 |
| Transv. emittance, rms | 0.02 ÷ 0.08 $\mu m$ |
| SR damping time | 0.5s ($5\times10^6$ turns) |
| RF V, f, harmonic | 75 kV, 162.5 MHz, 21 |
| Synchrotron tune | 0.005 ÷ 0.01 |

The goal of experiments at IOTA is to demonstrate the possibility to implement nonlinear integrable system in a realistic accelerator design. The project will concentrate on the scientific aspect of the single-particle motion stability in the nonlinear integrable system, leaving the studies of collective effects and attainment of high beam current to future research [7]. We intend to achieve the amplitude-dependent nonlinear tune shift exceeding 0.25 without degradation of dynamic aperture.

## NONLINEAR LENSES

Two classes of nonlinear focusing elements can be considered for the practical implementation: (1) a charge column (an electron lens) type and (2) an external static field type. The later is most restrictive because the static potentials in vacuum have to satisfy the Laplace equation.

An electron lens [6] employs the space charge forces of a low-energy beam of electrons that interacts with the high-energy particle over an extended length, $L_e$. The lens can be used as both linear and nonlinear focusing element depending on the electron current-density distribution $j_e(r)$ and the electron-beam radius $a_e$.

The first example of this type is the so-called McMillan thin lens. A 1-dimensional lens of this type was first described by E. McMillan [8] and was later extended into 2-d by Danilov and Perevedentsev [9]. For a thin lens approximation to be valid, the following condition must be met: $L_e<\beta$, where $\beta$ is the beta-function at the electron lens location. The electron lens current density has the following distribution:

$$j_e(r) \propto \frac{I_e}{\left(r^2 + a_e^2\right)^2} . \quad (4)$$

In this lens approximation such a charge distribution provides the following angular kick to a particle, passing through the electron lens at radius $r$:

$$\delta r' = \frac{kr}{r^2 + a_e^2} . \quad (5)$$

In order for this nonlinear system to be integrable, the IOTA ring must have the following one-revolution linear transformation matrix:

$$\begin{pmatrix} cI & sI \\ -sI & cI \end{pmatrix} \begin{pmatrix} 0 & \beta & 0 & 0 \\ -\frac{1}{\beta} & 0 & 0 & 0 \\ 0 & 0 & 0 & \beta \\ 0 & 0 & -\frac{1}{\beta} & 0 \end{pmatrix} \quad (6)$$

where

$$c = \cos(\phi),\ s = \sin(\phi),\ I = \begin{pmatrix} 1 & 0 \\ 0 & 1 \end{pmatrix} \quad (7)$$

and $\phi$ is an arbitrary parameter. Figure 2 shows an example of the tune footprint, obtained by the Frequency Map Analysis [10]. The attainable maximum spread of betatron frequencies for a single electron lens is ~0.3.

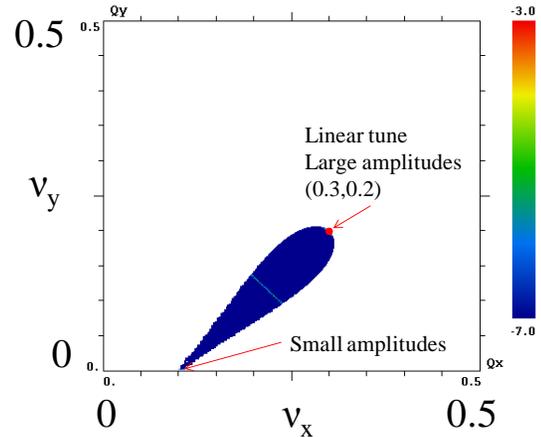

Figure 2: The fractional tune footprint in a thin-lens approximation ($L<<\beta$). For this simulation the 150-MeV IOTA beam was assumed to counter-propagate the electron lens beam current of $I_e \times L = 0.5$ A-m with $a_e = 1$ mm. The description of FMA simulation techniques can be found in Ref. [10].

The second example of a nonlinear system employing an electron lens does not require a thin-lens approximation. However, it requires an axially-symmetric electron lens current density distribution. We have used the distribution (4) in our simulations. The idea is based on the following principle: the electron lens guiding solenoidal field (5 kG) is sufficiently high to focus the

150-MeV IOTA beam as well. An example of ring beta-functions is shown in Figure 3.

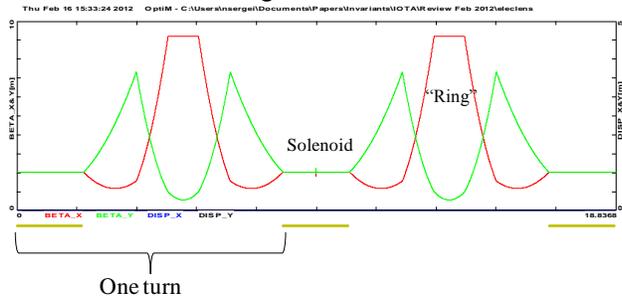

Figure 3: An example of "ring" beta-functions matched to a solenoid (5-kG) to create a section of constant beta-functions (2-m) within the solenoid length.

Figure 3 shows a very simplified "ring" functions. In reality, the only constraint on the optics function is that the ring (except for the solenoid) must have the betatron phase advance of $n\pi$, where $n$ is an integer, which is easily achievable in IOTA. Figure 4 shows the tune footprint for such a system.

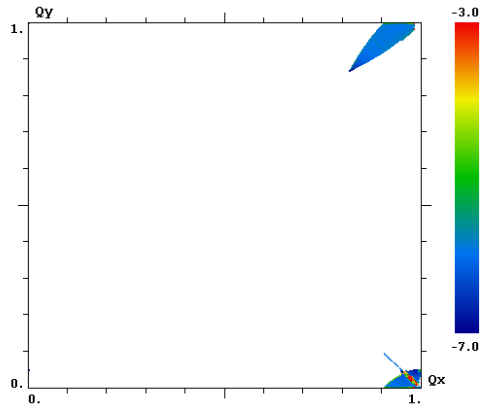

Figure 4: The fractional tune footprint for the case of an axially-symmetric electron lens (0.5 A-m) and a constant beta-function of 2 m. The achievable tune spread is ~ 0.25.

The final example of a nonlinear focusing system is based on a static magnetic field formed by external electromagnets. The optical principles for this case are described elsewhere [5]. The nonlinear magnet is 2-m long and consists of 20 short sections (Fig. 5).

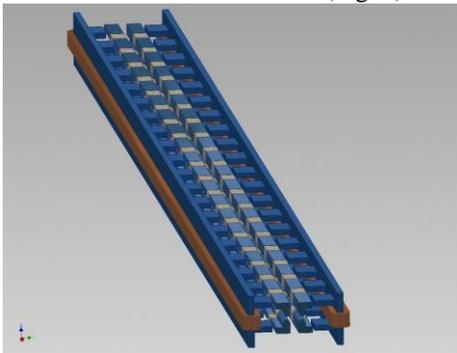

Figure 5: Conceptual design of a 2-m long nonlinear magnet for the IOTA ring.

The cross-section of the magnet is schematically presented in Fig. 6. Figure 7 presents the distribution of the main magnet parameters along its length. Figure 8 shows the tune footprint for IOTA with a single 2-m long magnet.

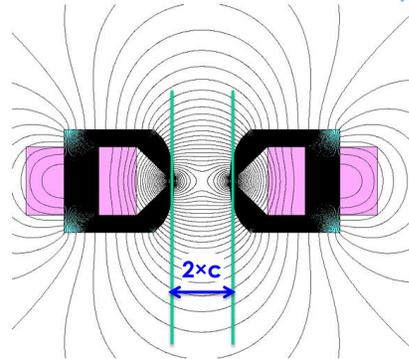

Figure 6: Cross-section of nonlinear magnet. The lowest order multipole expansion of this magnet is a quadrupole.

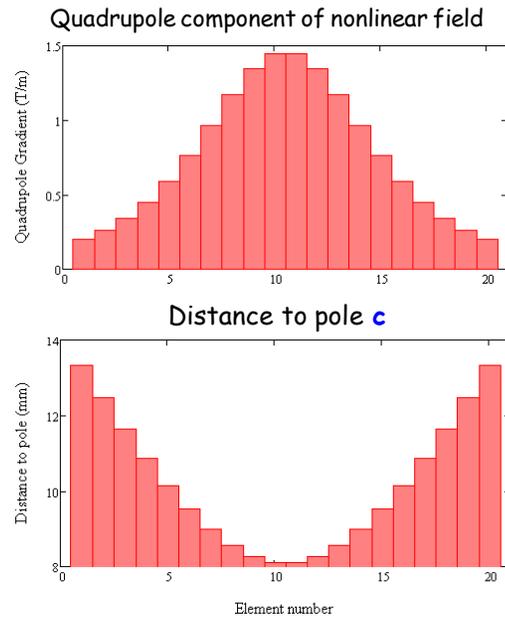

Figure 7: Quadrupole component and aperture parameter, $c$, along the magnet.

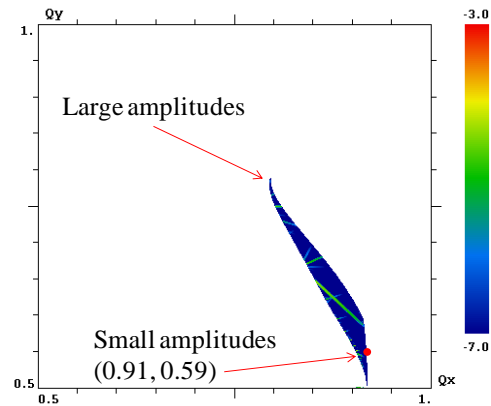

Figure 8: The IOTA fractional tune footprint for a single nonlinear magnet. The tune spread of ~0.3 is attainable.

## SUMMARY

In this paper we have presented first practical examples of completely integrable non-linear beam optics. These and other examples were modelled with single-particle tracking simulations [11] and demonstrated the viability of this concept. The Integrable Optics Test Accelerator (IOTA) ring is now under construction with the completion expected in 2014.

The ring can also accommodate other Advanced Accelerator R&D experiments and/or users. Current design accommodates Optical Stochastic Cooling

## ACKNOWLEDGMENTS

The authors thank V. Kashikhin for discussions on the design of nonlinear magnets, J. Leibfritz and L. Nobrega for help with the ring design. We are grateful to V. Shiltsev for advice, fruitful discussions, and his continuing support of this activity.

This work is supported by the U.S. Department of Energy under contracts No. DE-AC02-07CH11359. and No. DE-AC05-00OR22725.